\documentclass[aps, prl, twocolumn, showpacs, 10pt]{revtex4-1}
\usepackage{graphicx, amsmath, amssymb, bm}

\let\emptyset\varnothing

\begin{document}

\title{Noise-induced metastability in biochemical networks}

\author{Tommaso Biancalani}
%\email{tommaso.biancalani@gmail.com}
\author{Tim Rogers}
%\email{tim.rogers@manchester.ac.uk}
\author{Alan J.~McKane}
%\email{alan.mckane@manchester.ac.uk}

\affiliation{Theoretical Physics Division, School of Physics and Astronomy, 
University of Manchester, Manchester M13 9PL, United Kingdom}

\begin{abstract}
Intra-cellular biochemical reactions exhibit a rich dynamical phenomenology which cannot be explained within the framework of mean-field rate equations and additive noise. Here, we show that the presence of metastable states and radically different timescales are general features of a broad class of autocatalytic reaction networks, and that this fact may be exploited to gain analytical results. The latter point is demonstrated by a treatment of the paradigmatic Togashi-Kaneko reaction, which has resisted theoretical analysis for the last decade.
\end{abstract}


\maketitle

With recent advances in experimental techniques, it is becoming increasingly clear that the dynamics of cellular biochemical reactions are subject to a great deal of noise~\cite{Raj2009}. This poses a significant challenge to our understanding of such systems, as it has been known for some time that the effects of noise may lead to substantial differences in the macroscopic behavior~\cite{Rao2002,Maheshri2007}. The reactions which take place within a cell are highly interdependent, together forming biochemical networks which support the functioning of the cell. It remains a major open problem to make clear the link between the structural features of these networks and the resulting dynamics. A full understanding of the effects of noise is essential to this effort~\cite{Kaern2005,Shahrezaei2008}. 

Here, we report analytical progress on this problem made by studying a simple class of autocatalytic reaction networks whose dynamical behavior is radically affected by intrinsic stochasticity in finite volume cells. In particular, we show how networks of this type give rise to a separation of timescales between fast almost-deterministic oscillations and slow stochastic metastability. Our class includes the influential Togashi-Kaneko (TK) reaction scheme, numerical simulations of which have been found to undergo a noise-induced dynamical transition~\cite{Togashi2001,*Togashi2003}. Despite the importance of their work, a satisfactory analytic treatment of this effect has not been achieved in over a decade. Here we provide such a treatment as an application of our theory. 

The general model we work with is composed of $n$ chemical species, denoted by $X_i$ with $i=1,\ldots,n$, residing in a cell of (non-dimensional) volume $V$. The molecules undergo autocatalytic reactions of the form $X_i + X_j \rightarrow 2 X_j$, with rate coefficients $r_{ij}$. We put $r_{ij}=0$ if that particular reaction is not possible. We also stipulate that the total rates of creation and destruction of each reactant $i$ are in balance, that is, $\sum_jr_{ij}=\sum_jr_{ji}$. Two additional reactions, $\emptyset \rightarrow X_i$ and $X_i \rightarrow \emptyset$, represent diffusion into and out of the cell, respectively. The rate of diffusion is slow compared to the internal reactions, having coefficient $D\ll1$. We will also use the symbol $X_i$ to denote the number of molecules of that type, and $\bm{x}$ to indicate the concentration vector with components $x_{i}=X_i/V$.

The dynamics of the system defined by the above reactions are specified once the transition rates, $T(\bm{x} | \bm{x}')$, indicating the probability per unit of time that the system goes from state $\bm{x}'$ to state $\bm{x}$, are given. They are found by invoking mass action:
\begin{equation}
\begin{split}
&T\Big(x_i-\frac{1}{V}, x_j+\frac{1}{V}\,\Big|\, x_i, x_j\Big) = Vr_{ij}x_ix_j\,,\\
&T\Big(x_i-\frac{1}{V}\,\Big|\, x_i\Big) =DVx_i\,,\quad T\Big(x_i+\frac{1}{V}\,\Big|\, x_i\Big) =DV \,.
\end{split}\label{trates}
\end{equation}
The probability of finding the system in the state $\bm{x}$ at time $t$, $P({\bm{x}},t)$, then satisfies the master equation
\begin{equation}
\frac{dP({\bm{x}},t)}{dt} = \sum_{{\bm{x}}'\neq{\bm{x}}}\big[ T(\bm{x} | \bm{x}')P({\bm{x}}',t) - T(\bm{x}' | \bm{x})P({\bm{x}},t)\big],
\label{master} 
\end{equation}
with the transition rates given above \cite{Gardiner1985}. 

\begin{figure}
\begin{center}
\includegraphics[width=0.4\textwidth, trim=45 10 40 5]{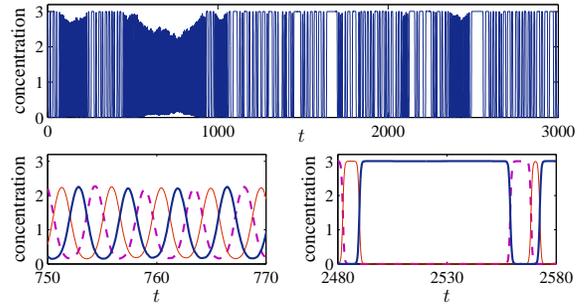} \\
\caption{(Color online) Sample stochastic time series of the simple three-species reaction network described in the text, with volume $V=10^4$ and diffusion coefficient $D=10^{-4}$. The thick (blue), thin (red) and dashed (purple) lines show the concentrations of chemicals $X_1$, $X_2$ and $X_3$, respectively. The smaller figures show detail of rapid oscillations (left) and metastability (right), taken from the main plot. All simulations were performed using Gillespie's algorithm \cite{Gillespie1977}.}
\label{fig:trajectories}
\end{center}
\end{figure}

Stochastic simulations of reaction networks of the class described above display a rich phenomenology including rapid oscillations and random switching between metastable states. For example, the time series displayed in Fig.~\ref{fig:trajectories} were obtained from simulations of a three-species reaction with (arbitrarily chosen) non-zero reaction rates $r_{1,2}=1$, $r_{2,3}=4$, $r_{3,2}=3$, $r_{3,1}=1$. In what follows we will show how these features can be qualitatively and quantitatively understood by an analysis of the influence of noise and the separation of timescales.

The dynamics are drastically affected by the relationship between the cell volume and the diffusion coefficient. To elucidate this, we introduce a rescaled volume $\lambda=DV$, which we treat as an $O(1)$ control parameter. Scaling $V$ and $D$ simultaneously in this way, we can rewrite the master equation (\ref{master}) as a power series in a single small parameter (we choose $D$, but $V^{-1}$ is also a valid expansion parameter), leading to a Kramers-Moyal expansion~\cite{Risken1989}. Truncating it at second order, one obtains a Fokker-Planck equation equivalent to the following stochastic differential equation (SDE), defined in the It\={o} sense~\cite{Gardiner1985}:

\begin{equation}
\label{sdes}
\dot x_i = x_i\sum_jR_{ij}x_j + D (1 - x_i) + \sqrt{D}\,\eta_i(t)\,,
\end{equation}
where $i=1,\ldots, n$, $R_{ij}=r_{ji}-r_{ij}$ and the $\eta_i$ are Gaussian noise variables with zero mean and correlator
\begin{equation}
\begin{split}
\big\langle {\eta_i}(t)& {\eta_j}(t') \big\rangle = \\
&\quad \delta(t-t')\,\frac{1}{\lambda}\left[  \delta_{i,j} \Big( x_i\sum_kS_{ik}x_k \Big)-S_{ij}x_i x_j \right].
\end{split}
\label{noise}
\end{equation}
Here the angle brackets signify an average over the noise, and $S_{ij}=r_{ij}+r_{ji}$.

Several important facts about the dynamics can be ascertained from inspection of Eqs.~\eqref{sdes} and \eqref{noise}. First, we discuss the limit of large volume. The factor of $\lambda^{-1}$ in the noise correlator indicates that for finite volumes the system experiences internal fluctuations. These vanish as $\lambda\rightarrow\infty$, leaving behind a deterministic system of differential equations equivalent to those obtained from a mean-field analysis of the reaction network. For general reaction networks these equations describe simple oscillatory relaxation towards the homogeneous fixed point $x_i=1$ for all $i$. This prediction is quite at odds with the rich phenomenology which is observed in stochastic simulations (as seen in Fig.~\ref{fig:trajectories}, for example). A proper treatment of the noise is thus necessary: from now on we keep $\lambda$ fixed and finite.

The presence of the small parameter $D$ in Eq.~\eqref{sdes} implies a separation of timescales. On an $O(1)$ timescale (which we refer to as {\it fast}), diffusion is negligible and the system feels no noise. Setting $D=0$ in Eq.~\eqref{sdes} yields a deterministic dynamical system in which the homogeneous state $x_i\equiv1$ is a center; it has Jacobian matrix $R$, which is antisymmetric and thus has all imaginary eigenvalues. We can therefore expect rapid almost-deterministic oscillations as seen, for example, in the lower left panel of Fig.~\ref{fig:trajectories}. On a slow $O(1/D)$ timescale, two additional factors play a role. First, the system experiences a deterministic linear drag towards the homogeneous state. Second, the effects of noise become relevant, leading to stochasticity in the trajectories. 

For smaller volumes, the overall noise strength is greater, and thus the form of the noise correlator \eqref{noise} has an important role in shaping the system dynamics. In particular, since the strength of the noise is a function of the state of the system, trajectories are forced away from states giving rise to large values of noise, creating an effective attraction towards those states in which the noise vanishes. This effect is relatively well-known in the study of systems with multiplicative noise (for example, see \cite{Horsthemke1984} and references therein), and we will illustrate it with an explicit calculation for the TK model. Inspection of the correlator \eqref{noise} reveals that the states for which the noise vanishes are those in which no autocatalytic reaction can occur. That is, for each pair $i,j$ one of $x_i$, $x_j$ or $r_{ij}$ must be zero. The metastability of these states is further enhanced by the fact that this condition also causes the $O(1)$ term in Eq.~\eqref{sdes} to vanish. An example can be seen in the lower right panel of Fig.~\ref{fig:trajectories}, where the state $X_1=3,\,X_2=0\,,X_3=0$ is metastable. 

As well as providing a qualitative picture of dynamics observed in this class of biochemical reaction networks, the mathematics we describe may also be employed to obtain precise analytical results \footnote{Timescale separation techniques have also been applied successfully to other models with intrinsic noise, for example, in \cite{Parker2009} to study the properties of stochastic extinction events in the Lotka-Volterra model}. We now illustrate these methods in the paradigmatic case of the TK reaction~\cite{Togashi2001,*Togashi2003}. The model is composed of four chemical species whose reactions form a closed cycle, so that the non-zero rates are $r_{1,2}=r_{2,3}=r_{3,4}=r_{4,1}=1$. In stochastic simulations of the model, different dynamics are observed depending on the volume of the cell. For very large volumes, one finds an approximately homogeneous distribution of chemical species, however, at lower volumes the system is typically dominated by a pair of species (either $X_1$ and $X_3$, or $X_2$ and $X_4$), with the other pair absent: these are the metastable states predicted in the earlier discussion.

\begin{figure}
\begin{center}
\includegraphics[width=0.4\textwidth, trim=15 20 40 6]{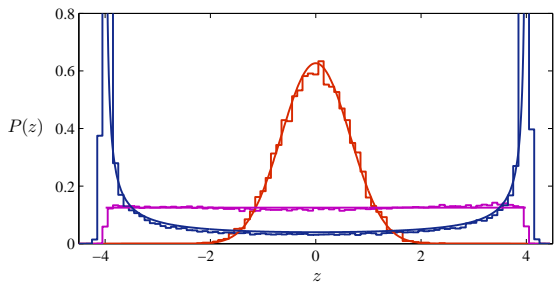} \\
\caption{(Color online) Stationary probability distribution for $z=(x_1+x_3) - (x_2+x_4)$ in the TK reaction. The histograms were obtained from simulation data with diffusion coefficient $D=5\times10^{-3}$ at volumes $V=10^{4}$ (red, unimodal), $V=2\times10^{3}$ (purple, flat) and $V=10^{3}$ (blue, bimodal). In each case the corresponding theoretical prediction of Eq.~\eqref{zstaz} is shown with a solid line.}
\label{fig:trans}
\end{center}
\end{figure}

To visualize this dynamical transition, TK~\cite{Togashi2001,*Togashi2003} introduced the quantity $z = (x_1+x_3) - (x_2+x_4)$. The pair-dominated state corresponds to $|z|\approx 4$. By measuring the stationary distribution $P(z)$ from long simulation runs, one observes a transition induced by cell volume -- see Fig.~\ref{fig:trans}. There is a critical volume $V_c\approx 1/D$ at which $P(z)$ is flat; above $V_c$ the distribution has a single peak at $z=0$; below $V_c$ it is bimodal with peaks at $z\approx\pm 4$, indicative of the pair-dominated regime. 

In large volumes the model also exhibits quasi-cycles, a second (weaker) stochastic effect whereby damped oscillations present in the deterministic dynamics are excited by the noise. Quasi-cycles are amenable to analysis using a linear noise approximation \cite{Dauxois2009}, however, it is clear that the dynamical transition is related to the noise-induced metastability discussed above and will require more powerful methods. This point was elucidated in Ohkubo {\it et al}~\cite{Ohkubo2008}, with the investigation of a simple one-dimensional model inspired by the TK reaction. 

We begin our analysis of Eq.~(\ref{sdes}) for the TK reaction by making a change of variables which can be understood mathematically (as a real Fourier transform) or physically (as corresponding to the total concentration, the $z$ variable introduced by TK and two variables related to the $X_{1}-X_{3}$ and $X_{2}-X_{4}$ dynamics). This is
\begin{eqnarray}
& & w=x_1 + x_2 + x_3 + x_4, \  z=(x_1+x_3)-(x_2+x_4), \nonumber \\
& & u=x_1-x_3, \ v=x_2-x_4.
\label{CofV}
\end{eqnarray}
Applying the transformation, for the total concentration we find the closed equation $\dot{w}=D(4-w)$. For the remainder of the analysis, we fix $w$ to its fixed-point value of $4$. For the variables $z$, $u$, and $v$ we then find
\begin{equation}
\begin{split}
  \dot z &= -2 u v- D \,z + \sqrt{\frac{D}{\lambda}(16-z^2)}\,\zeta_1(t)\,\,, \\
  \dot u &= - \frac{v(z+4)}{2} - D\,u +\sqrt{\frac{D\,(4-z)}{\lambda\,(4+z)}}\,\Big(u\,\zeta_1(t)+\phi\,\zeta_2(t)\Big)\,, \\
  \dot v &= - \frac{u(z-4)}{2} - D\,v -\sqrt{\frac{D\,(4+z)}{\lambda\,(4-z)}}\,\Big(v\,\zeta_1(t)+\psi\,\zeta_3(t)\Big) \,,
\end{split}
\label{sdes2}
\end{equation}
where $\phi = \sqrt{(z + 4)^2/4-u^2}\,$, $\psi=\sqrt{(z - 4)^2/4-v^2}$, and the $\zeta$ variables are independent Gaussian white noise.\par
The dynamics on the $O(1)$ timescale are solvable. In fact, $\phi$ and $\psi$ defined above are conserved quantities of the system \eqref{sdes2} with $D$ set to zero. Solution trajectories are therefore confined to the closed curve given by the intersection of the surfaces defined by the values of $\phi$ and $\psi$, which are determined by initial conditions. Details of the full solution will be provided in a forthcoming paper \cite{Biancalani2012}; for the present discussion it is sufficient to point out that the trajectories are periodic, with the period for $z$ being 
\begin{equation}
\label{o1period}
T = \frac{2}{\sqrt{16- \left( \phi^2 - \psi^2 \right)}}\,\, \text{K} \left( \frac{16 - (\phi + \psi)^2}{16 - (\phi - \psi)^2} \right),
\end{equation} 
where $\text{K}(\cdots)$ denotes the elliptic integral of the first kind. The period for $u$ and $v$ is double that of $z$. It is important to note that $K(x)$ grows without bound as $x\to1$, and thus Eq.~(\ref{o1period}) implies that the period of oscillation $T$ diverges as either $\phi\to0$ or $\psi\to0$. In these limits, the trajectories of the deterministic dynamics deform into a homoclinic network linking the fixed point $(u,v,z)=(0,4,-4)\,\,\textrm{to}\,\,(u,v,z)=(0,-4,-4)$ or $(-4,0,4)\,\,\textrm{to}\,\,(4,0,4)$, respectively. This fact explains the presence of both fast oscillatory dynamics and metastability in the same parameter range.

We turn now to the study of the behavior of $z$ on an $O(1/D)$ timescale. From left to right, the terms in the equation for $\dot{z}$ in system \eqref{sdes2} are responsible for the fast oscillation caused by interaction with $u$ and $v$, the linear drag towards zero, and the noise. Since the oscillations occur on a timescale faster than the other two terms, we expect that a time average on a timescale $\tau$, such that $T \ll \tau \ll 1/D$, will not affect the drag and the noise substantially. To do the averaging, we coarse-grain time by intervals with length $\tau$ in Eq. \eqref{sdes2} and replace every term with its time average over that interval \cite{Freidlin1984}. We write $\overline{(\cdots)} = \tau^{-1} \int_t^{t+\tau}dt \,(\cdots)$ for the time average and make use of the following assumptions:
\begin{equation}
\overline{\vphantom{i} u v}\,\approx0\,,\quad \overline{\left(16-z^{2}\right)^\frac{1}{2}} \,\approx\,\left(16-\overline{z}^2\right)^\frac{1}{2}\,.
\label{ansatz}
\end{equation}
These are justified on physical grounds: the first follows from the fact that the conserved quantities of the fast dynamics are approximately constant on intervals of length $\tau\ll 1/D$, since the average of $uv$ is a multiple of the average of $\dot{z}$, and $z$ has periodic trajectories if $\phi$ and $\psi$ are fixed; in the second approximation, we are assuming that the strength of noise is not strongly affected by fast oscillations in $z$. 

The resulting so-called averaged equation for $\bar z$ is

\begin{equation}
\dot{\bar{z}} = - D \bar{z} + \sqrt{\frac{D}{\lambda}\,(16-\bar z^2)}\,\zeta(t)\,.
\label{zeq4}
\end{equation}
This equation describes an interplay between the drag and the noise, and provides a complete picture of the dynamical transition first observed by TK. Physically, we may think of the system as gently relaxing to the origin, while being agitated by a noise term which vanishes at the metastable states $\bar z=\pm4$. Depending on the strength of the noise (controlled by the parameter $\lambda$), the system will either be attracted to zero by the linear drag, or forced to the boundaries by the noise. By varying $\lambda$ we transition between these dynamical regimes, an effect which is most clearly demonstrated by calculation of the stationary distribution $P(\bar z\,;\,\lambda)$. From Eq.~\eqref{zeq4}, we find

\begin{equation}
P(\bar z\,;\,\lambda) = \big(16-\bar z^2\big)^{\lambda -1} \frac{\Gamma\left(\frac{1}{2}+\lambda \right)}{\sqrt{\pi }\, 4^{2 \lambda-1}\, \Gamma(\lambda)}.
\label{zstaz}
\end{equation}

Our prediction is tested against the numerics in Fig.~\ref{fig:trans}. This equation confirms the critical volume $V_c=1/D$ as the point of transition between a unimodal and bimodal stationary distribution. We should point out that Eq.~\eqref{zstaz} is correct only up to first order in $D$; certain features of the simulation data (such as $|z|$ occasionally exceeding 4 due to variations in total concentration $w$) are not captured at this level of approximation.  

It is worth pausing a moment to discuss the relation of the noise-induced metastable states to the fast oscillatory dynamics discussed in the earlier analysis. For example, from the definition of $\psi$, we see that $z=4$ can only be obtained when $v=0$ and $\psi=0$, and thus we are in the regime in which the period of the oscillation is divergent. In this case one can expect fast periodicity to break down and the system to remain in a given metastable state for a random length of time, before being freed and proceeding along a trajectory close to the homoclinic orbit linking it to another. The lower right-hand plot of Fig.~1 shows this behavior.

Beyond determining the stationary distribution of $z$, our methods may also be used to calculate various other quantities associated with the model. For example, in \cite{Togashi2001} the fraction of time spent in the pair-dominated state (that is, $X_1+X_3=0$ or $X_2+X_4=0$), called the `rate of residence', was measured from simulations and plotted as a function of $\lambda$. The authors noted a puzzling shift in this quantity when adjusting for different cell volumes, which we are now able to explain.

\begin{figure}
\begin{center}
\includegraphics[width=0.4\textwidth, trim=25 20 40 3]{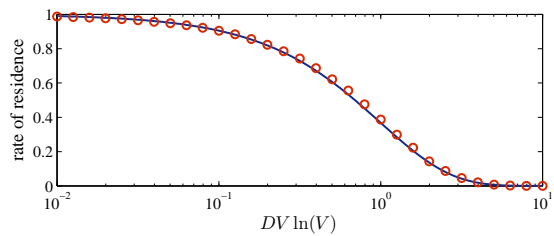} \\
\caption{(Color online) Rate of residence of the pair-dominated state as a function of $DV\ln(V)$. The circles show the result measured from simulations carried out with fixed $D=10^{-3}$ and varying $V$; for each data point a single simulation of duration $t_{\max}=10^7$ was conducted and the fraction of time spent in the pair-dominated state measured. The solid line corresponds to the first-order prediction in Eq.~\eqref{dvlogv}.}
\label{fig:rateofres}
\end{center}
\end{figure}

From Eq.~\eqref{zeq4} we can determine a straightforward prediction for the rate of residence by computing the fraction of time that $z$ spends within $1/V$ of $\pm4$. We integrate the stationary distribution to find:
\begin{equation}
\begin{split}
1-\int_{-4+1/V}^{4-1/V}P(z\,&;\,DV)\,dz\\
&=\Big.V^{-DV}+\text{higher order terms.}
\end{split}\label{dvlogv}
\end{equation}
Therefore, to properly compare different cell volumes and diffusion coefficients, one should hold $DV\ln(V)$ constant, rather than $\lambda$. The fit between Eq.~\eqref{dvlogv} and data from simulations is shown in Fig.~\ref{fig:rateofres}. 

In this Rapid Communication we have examined the influence of noise on the link between structure and function in a class of biochemical networks.  The consistent formulation of the problem which we provide starts from the master equation and proceeds through a well-defined approximation scheme to an SDE which correctly captures the behavior of the system. Although this equation is not exactly solvable, we are able to proceed by identifying and exploiting a separation of timescales involved in the problem. This analytical process was demonstrated explicitly for the paradigmatic TK reaction, providing an understanding of the phenomenology of the model and yielding expressions for quantities of interest which are compared to the ones obtained numerically by TK.

Since it is the discreteness of molecules which gives rise to the intrinsic noise experienced by reaction systems of this type, one might expect that such effects are only relevant in small systems, and can be neglected in general (indeed, this is a central assumption of any theory based on the study of macroscopic rate equations). In practice the situation is far more subtle; what matters more than the strength of the noise is how it interacts with other aspects of the model, such as the slow relaxation due to a small diffusion coefficient. As we have shown, this interaction gives rise to metastability in the class of autocatalytic reaction networks we investigate; moreover, it can be exploited mathematically to explain the dynamical transition observed in the TK reaction. A closely related noise effect has recently been observed in an ecological model \cite{Rogers2012}, where it induces the spontaneous formation of species, and we expect that more surprising results of this type will come in the near future.

\vspace{10pt}
This work was funded (T.R. and A.J.M.) under EPSRC Grant No. EP/H02171X/1. T.B. also acknowledges partial funding from EPSRC.

\end{document}